\documentstyle[12pt]{article}
\title{On the environmental decoherence and spin interference in mesoscopic loop structures}
\author{I.Tralle \footnote{
Corresponding author e-mail addresses: tralle@univ.rzeszow.pl},
$\;$ W. Pa\'sko \\
{\em Institute of Physics, University of Rzesz\'ow}\\\em
Al. T. Rejtana 16A, 35-310 Rzesz\'ow, Poland}
\date{}

\begin{document}
\maketitle
\begin{abstract}

    Mechanisms of 'environmental decoherence' such as surface scattering, Elliot-Yafet process and precession mechanisms, as 
well as their influence on the spin phase relaxation are considered and compared. It is shown that the 'spin ballistic' 
regime is possible, when the phase relaxation length for the spin part of the wave function $(L_{\varphi}^{(s)})$ is much 
greater than the phase relaxation length for the 'orbital part' $(L_{\varphi}^{(e)})$. In the presence of an additional 
magnetic
field, the spin part of the electron's wave function (WF) acquires
a phase shift due to additional spin precession about that field.
If the structure length $L$ is chosen to be $L_{\varphi}^{(s)} > L
> L_{\varphi}^{(e)}$, it is possible to 'wash out' the quantum
interference related to the phase coherence of the 'orbital part'
of the WF, retaining at the same time that related to the phase
coherence  of the spin part and, hence, to reveal corresponding
conductance oscillations.

\end{abstract}

\section{Introduction}

One of the main ideas which underpins different possible applications of 'spin transport',
including information storage and computation, is that the spins of electrons in semiconductors
may have very long quantum coherence times [1,2], or in other words, electrons can travel a long
way without flipping their spins. But this also gives the possibility to observe quantum effects
which involve the interference of electron waves. In the classical picture of transport phenomena,
the total probability for a particle to transfer from one point to another is the sum of the
probabilities for such transfer over all possible trajectories. In the quantum description, this
result corresponds to neglecting the interference of scattered electron waves propagating along
different paths. The destruction of quantum coherence is controlled by the phase relaxation time
or phase relaxation length. Since for the electron spin this length may be very long, it is naturally
to expect that the spin interference can reveal itself in the conductance oscillations similar to that
ones which are due to Aharonov-Bohm effect [3]. Most of the researchers who dealt with the
Aharonov-Bohm effect considered mainly the Hamiltonian
\begin{math}\hat{H} = (\bf p \rm - (e/c) \bf A\rm)^{2}/2m^{*} + U(y,z)\end{math}, where U(y, z)
is the energy corresponding to the transverse motion, and almost nobody takes into account the spin-part
 \begin{math}\mu_B\hat{\sigma}\bf B\end{math} of the
Hamiltonian ($\mu_B$  is the Bohr magneton,
\begin{math}\hat{\sigma}\end{math} is the electron spin operator,
\bf B \rm - magnetic field). The reason for the neglecting of this
part of the Hamiltonian is that one usually treats  electron's spin as something which is of no importance for the transport 
properties (for instance, current) of solids, semiconductors in particular. However, if the quantum interference is 
concerned, it might happen that the spin of electrons can play an important role [1,2,4].  
 If one considers the total
Hamiltonian which includes Pauli term, one inclines (being in the framework of Pauli's spin theory) to treat $\sigma=\pm 
\hbar /2$ as 'the projection of spin vector onto quantization axis' which coincide with external magnetic field $\bf B$. But 
everybody who ever dealt with Dirac equation knows pretty well that it is not the case. May be it is worth-while to quote 
here the words of one of the most influential contemporary mathematicians, Yu. Manin, who wrote [5]: '... we only mislead 
ourselves if we awkwardly try to describe the internal quantum degrees of freedom as the 'value of projection of spin on the 
$z$-axis', since a spin vector lies in a completely different space from the $z$-axis'. Nevertheless,this na\"\i{}ve 
treatment of $\sigma$ as the 'projection of spin vector onto $z$-axis', to paraphrase John Bell's words, is sufficient for 
\it almost \rm  all practical purposes. Moreover, it can be easily amended by taking into account the true structure of 
corresponding Hilbert space and simply by writing down the electron wave function in factorized form as the direct product:
$\Psi({\bf r},s) = \varphi({\bf r})\otimes \chi(s)$. It is worthy to note also, that if the quantum interference is 
concerned, the quantity of main importance is the phase coherence length. Introducing the factorized form of the wave 
function, one not only can, but rather have to introduce simultaneously two phase  relaxation lengths, the first one for the
'orbital part' of the electron wave function, $L_{\varphi}^{(e)}$,
and the second one, $L_{\varphi}^{(s)}$ for the spin part. It turns out [4], that $L_{\varphi}^{(s)} >>
L_{\varphi}^{(e)}$, which is in total agreement with the experiment
[1,2,6]. The physics which is behind that, is the following. An
electron during its transfer along some path in the solid
(semiconductor, for definiteness) interacts all the time with the
environment. As a rule rigid scatterers such as impurities and
other defects of crystalline structure, do not contribute to the
phase relaxation; only dynamical scatterers like phonons, do. On
the other hand, the electron scattering by phonons is mainly
inelastic, while impurity scattering is mainly elastic, so we can
say that only inelastic scattering contributes to the phase
relaxation. But what does it mean inelastic scattering in case of
spin? It means spin flips which are the consequences of
scattering by phonons accompanied by spin-orbit interaction. This
interaction is  weak and that is why the spin flips are rare
events and the phase relaxation length for the spin part of the
electron wave function is very long. But now, if the structure
length $L$ is chosen to be $L^{(s)}_{\varphi} > L >
L^{(e)}_{\varphi}$, it is possible to 'wash out' the quantum
interference related to phase coherence of the 'orbital part' of
the wave function retaining at the same time the phase coherence
of the spin part one and hence, to reveal the corresponding
conductance oscillations of the microstructure. Such model was
considered in the  paper [4] where the simple theory of the
quantum interference in a loop structure due to Larmor precession
of electron spin was presented for
the first time. The aim of this paper is to develop the approach
further, discussing other aspects of the problem such as, for
instance, electron scattering by the edges of the structure, the 'precession mechanisms' of spin relaxation which are 
characteristic ones for the crystal with lack of inversion symmetry, as well as their influence on the quantum interference 
of spins.

\section{Scattering by the edges of the structure}

   We start with a generic microstructure with two end regions (\begin{math} x < 0 \end{math} and \begin{math} x >L\end
{math})
and a middle region \begin{math}0 \leq x\leq L\end{math}
consisting of two channels (Fig. 1),very similar to that one considered
in [4]. The main difference however is the following. In the Ref
[4], apart an external magnetic field $\bf B_{0}$ in the plane of
the microstructure, on the upper surface of one of the channels
there was a regular periodic array of micromagnets which created
and additional magnetic field. As it was mentioned in [4], the
periodic magnetic field was not obligatory and that choice was
motivated by the current interest in the study of electron motion
in inhomogeneous magnetic fields on the nanometer scale. Here
instead, even more simpler case of uniform magnetic field is
considered, since the only thing which is really needed is that
the magnetic fields to be different in the two arms of the loop.
An additional advantage of the uniform magnetic field is that it
is more simple and appropriate from the experimentalist's point of view.

Consider an electron enters the domain occupied by the magnetic
field, say, from the left-end region. The electron's spin wave
function is a coherent superposition of the spin-up and spin-down
eigenstates, which are split in the magnetic field by the Zeeman
energy $\Delta \varepsilon=g\mu_B B$, ($g$ is the Lande factor).
Coherent evolution under the spin Hamiltonian results in
oscillations between these two eigenstates; classically this
oscillation corresponds to precession of the spin vector at the
Larmor frequency $\Delta \varepsilon/\hbar$. In other words, we
consider the non-relativistic electron motion in the magnetic
field as the motion of a classical top which precesses about the
magnetic field. Since the magnetic fields are different in the two
arms of the structure, the phase shifts acquired by the spin wave
functions are also different and if one of the field (say, $B_0$)
 alters, it should lead to the specific conductance
oscillations dependent on $B$.

So, the magnetic field {\bf B} which affects the electron in the first arm of the structure is equal to ${\bf B}$= ${\bf B_0}
$,
while in the other one it is equal to ${\bf B}$=${\bf B_0}$+${\bf B_1}$.
Suppose the Hamiltonian of an electron is $H=H_0+H_1$, where
\begin{equation}
H_0=1/2m^{\ast}\left(
{\bf p} -\left(e/c\right){\bf A}\right)^2+U({\bf r})\;\;,
H_1=-\mu_B\hat{\sigma}\bf B\;\;,
\end{equation}
where $m^{\ast}$ is the electron effective mass, ${\bf A}$
is vector potential corresponding to the magnetic field ${\bf B}$, $\mu_B$ and $\hat{\sigma}$
are Bohr magneton and the spin operator respectively. We also assume that $U({\bf r})$
describes conduction bands bending due to space charge and discontinuities of any band. Since $H_0$
does not depend on spin, the wave function is the direct product: $\Psi({\bf r},s)$ = $\varphi({\bf r})$$\otimes$$\chi(s)$.
Ever since for convenience we shall refer to $\varphi({\bf r})$ as the 'orbital part' of the total wave function, keeping in 
mind
that it corresponds to $H_0$ describing the charge-field interaction, and we shall refer to $\chi(s)$ as the spin-part of the 
wave
function related to $H_1$, the spin part of the Hamiltonian $H$ in (1).

Let us now introduce  the phase-relaxation length
$L_{\varphi}^{(s)}$ for the spin part of the wave function, in
just the same way as the one usually introduced for the 'orbital
part', $L_{\varphi}^{(e)}$. As it was argued in the Introduction,
the phase relaxation length $L_{\varphi}^{(s)}$ is much greater
than $L_{\varphi}^{(e)}$. The proof of the statement can be found
in Ref.[4], here we would like only to emphasize, that in accordance with our very cautions estimates, the phase relaxation 
time for the spin part of the wave function, $t\sim\tau^{(s)}_{\varphi}$ is about $\sim 2.2\times 10^{-10} s$. This time is 
indeed much longer than the phase relaxation time for the orbital part one [7], and hence, if one takes into account only 
single scattering mechanism (scattering by phonons; this mechanism is known in the literature as Elliot-Yafet process [8]), 
one can safely assume $L^{(s)}_{\varphi}>>L^{(e)}_{\varphi}$ and consequently, that the structure length $L$ can be chosen to 
be $L_{\varphi}^{(s)}\;>\;L\;>\;L_{\varphi}^{(e)}$.

However, there is another mechanism which also, in principle, can contribute to the spin relaxation. This mechanism is the
scattering by the surface and the  edges of the structure. 

    Now our aim is to calculate the probability of the spin-flips caused by the electron scattering by the surface of the 
structure,
since such spin-flips results in the 'washing out' of 'phase memory' of the spin-part of the electron wave function.

It is commonly used to treat the interaction of electrons with the surface of the sample in terms of a phenomenological 
parameter
 $\epsilon_0$ introduced for the first time by F. Dyson [9]. This quantity can be defined as the mean probability of spin-
flip of
the conduction electrons having the energy ${\cal E}={\cal E}_F$ at their collision with the surface of the sample, averaged 
over the
incident angles. Following [9], define $\epsilon_0$ as
\begin{displaymath}
\epsilon_0=\hbar^{-1}\int d\Omega(\bf J\rm S)^{-1}\int d\bf k^{\prime}\rm \rho(\bf k^{\prime}\rm ) \delta [{\cal E}(\bf k \rm 
,\sigma) - {\cal E}
(\bf k^{\prime}\rm ,\sigma^{\prime})]|V_{\bf k^{\prime}\rm,\sigma^{\prime};\bf k\rm ,\sigma}|^{2},
\end{displaymath}
where $\Omega$ is the incident solid angle,
$\sigma,\sigma^{\prime}$ are the spin variables corresponding to
the states before and after the collision with the surface, $\bf k
\rm , \bf k^{\prime}$ \rm are the electron wave vectors before and
after collision; $\bf J \;$\rm is the flux density of electrons
incident on the surface from within; S is the surface area,
$\rho(\bf k^{\prime}\rm )$ is the density of states in the \bf k
\rm -space over the single spin; $V_{\bf
k^{\prime}\rm,\sigma^{\prime};\bf k\rm ,\sigma}$ is the matrix
element of the perturbation operator responsible for the
spin-flips; ${\cal E}(\bf k \rm ,\sigma), {\cal E}(\bf
k^{\prime}\rm ,\sigma^{\prime})$ are the electron energies in the
corresponding quantum states.

It is quite obvious that the crystal potential which is supposed to be periodic deep inside the sample, changes abruptly at 
the
edges and vicinity of the surface where it is not periodic altogether. These abrupt changes of the potential lead to the 
emerging
of an electric field in a thin layer near the surface of the sample. The thickness of the layer can be estimated as to be 2-4
monolayers. Hence, if the electrons are moving in this layer, one can treat it as the quasi-two dimensional electron gas 
(2DEG).

Introduce now the space variables ${x,y,z}$ where $z$-axis is normal to the 2DEG. Suppose the external magnetic field $\bf B 
$
is directed along $y$-axis, $\bf B \rm =(0, B, 0)$ and let the vector potential $\bf A$ \rm to be: $\bf A \rm = (Bz, 0, 0)$. 
Now
the Hamiltonian of the electron moving in our 2DEG is of the form:
\begin{equation}
H=\frac{1}{2m^{*}}
\left[
\left(
p_x - \frac{eBz}{c}
\right)
^{2} + p_y^{2} + p_z^{2}
\right]
 + U(z) + \mu_B\hat{\sigma}B_0. \end{equation}
Here $U(z)$ is the potential responsible for the space quantization in 2DEG in $z$-direction. The electron functions 
corresponding
to such Hamiltonian can be written down as \begin{equation}
\Psi_{\bf k \rm \sigma}(\bf r \rm )= C\exp(ik_x + ik_y)\varphi(k_x, z)\exp(-i{\cal E}_{k}/\hbar)\otimes \chi_{\sigma}\end
{equation}
and the normalizing conditions defined as
\begin{displaymath}
\int dx\psi^{*}_{k_x^{\prime}k_y^{\prime}\sigma^{\prime}}(x,y)\psi_{k_x k_y \sigma}(x,y)=\delta(\sigma,
\sigma^{\prime})\delta(\bf k\rm - \bf k^{\prime}\rm )\end{displaymath} \begin{displaymath}
C=((L_xL_y)\int^{\infty}_{-\infty}|\varphi(k_x,z)|^{2}dz)^{-1/2} .\end{displaymath}
Here $L_x,L_y$ are the structure sizes in $x,y$-directions.

As it was mentioned above, due to the potential $U(z)$, the strong electric field emerges at the surface of the structure . 
Obviously,
the field is equal to $E=-e^{-1}\frac{\partial}{\partial z}U(z)$. If we suppose the absence of magnetic
impurities in semiconductor, the only mechanism responsible for the spin-flips is the spin-orbit interaction. In the refrence 
frame
of moving electron, the electric field generates an effective magnetic field of the form $\sim [\bf E \rm \times (d\bf r \rm 
/dt)c^{-1}]$
which causes spin flips. The operator of spin-orbit interaction has the form (see Ref. [10]):
\begin{equation} V^{so} = \frac{\hbar^{2}}{4m^{*2}c^{2}}\frac{\partial}{\partial z}U(z)
(\sigma_y \frac{1}{i}\frac{\partial}{\partial x} - \sigma_x\frac{1}{i}\frac{\partial}{\partial y}), \end{equation}
where $\sigma_y, \sigma_x$ are the corresponding Pauli matrices and this is the perturbation mentioned above.

After some manipulations carried out much in the same way as in Ref.[11], the dimensionless probability of the spin
flips caused by the potential (4) and averaged over the incident angles, can be written as:
\begin{displaymath} \epsilon_0 = \frac{1}{60}(mc^{2})^{-2}(cos^{2}(\widehat{\vec{B},\vec{n}})+1)I, \end{displaymath}
where $(\widehat{\vec{B},\vec{n}})$ is the angle between $\bf B$ \rm and the unit vector $\vec{n}$ normal to the surface and 
\begin{equation}
I=|\int^{\infty}_{-\infty}dz\varphi^{*}(-k_x, z)\frac{\partial U}{\partial z}\varphi (k_x, z)|^{2}.\end{equation}

Assuming, as it was done in [11], that the parameter $b=
g\mu_B B/{\cal E}_F \ll 1$ ($g$ is the Lande factor), it is
possible to calculate integral in (5) even without having known
the wave functions $\varphi_{k_x z}$ and the form of the potential
$U(z)$. The authors of [11] considered the electron scattering by
the surface in a metal where the Fermi energy is high enough and
the parameter $b$ is small. However, in our case this
assumption is no longer valid, since in 2DEG Fermi energy is not
so high. So, we calculate the integral in (5) using the  simple
model of 'triangular potential well'. To this end, we write down
the Schr\"odinger equation for the 'orbital part' of the wave
functions (3) (it corresponds to the Hamiltonian (2) without
Pauli's term) in the form:
\begin{equation}
-\frac{\hbar^{2}}{2m^{*}}\frac{\partial^{2}\varphi(k_x,z)}{\partial z^{2}} + 
\left[
\frac{\hbar^{2}k^{2}_x}{2m^{*}} +
\frac{\hbar eBk_x z}{m^{*}c} + \frac{(eBz)^{2}}{2m^{*}c^{2}} + U(z)-{\cal E}(k_x,k_y)
\right]
\varphi(k_x,z) = 0, \end{equation}
where ${\cal E}(k_x,k_y) = {\cal E}_{k_x} - \hbar^{2}k^{2}_y/2m^{*}$.

Now suppose the potential $U(z)$ has the form of triangular potential well, that is
\begin{displaymath}
U(z)=\left\{
\begin{array}{cc}
\infty, & z < 0 \\
eEz, & z \geq 0
\end{array} \right. \end{displaymath}
Suppose also  so called quantum limit to occur, when only the lowest bands are occupied by the electrons. In the absence of a
magnetic field, electrons in lower subbands (small subband index $n$) have a large mean free path, because the amplitude near 
the
boundary is smaller and the scattering is less frequent. Therefore, the current is carried mostly by electrons in lower 
subband.
In a magnetic field, electrons are pushed toward the boundaries by a Lorentz force and the scattering is enhanced. From now 
on we
assume also the condition $(d_0/l_B)^{4}\ll 1$ is fulfilled. In the last relation $d_0 = (3\hbar^{2}\pi^{2}/16m^{*}eE)^{1/3}
$ is the
thickness of the 2DEG layer and $l_B = (\hbar c/eB)^{1/2}$ is the so called magnetic length. Then, the solution of the 
eigenvalue
problem for Eq.(6) in the quantum limit approximation is of the form (see [12]):
\begin{displaymath}
\varphi(k_x, z)=\left\{
\begin{array}{ll}
Ai([\frac{2m^{*}eE}{\hbar^{2}} + \frac{2eBk_x}{\hbar c}]^{1/3}[z - \frac{{\cal E}(k_x, k_y)-\hbar^{2}k^{2}_x/2m^{*}}{eE+\hbar 
eBk_x/m^{*}c}]),& z\geq 0\\
0,& z < 0 \end{array} \right. \end{displaymath}
\begin{displaymath} {\cal E}(k_x, k_y) -
\hbar^{2}k^{2}_x/2m^{*}=(\hbar^{2}/2m^{*})^{1/3}[\frac{9\pi}{8}(eE
+ \frac{\hbar eBk_x}{m^{*}c})]^{2/3}, \end{displaymath} where
$Ai(...)$ means Airy function (see, for instance, [13]). By means
of the last formulae, we evaluated numerically the integral in (5)
(see Fig. 2); at the numerical estimates of the spin flip
probability we used also the next values of the parameters: $B\sim
0.1 T, k_x\sim k_F=0.3\times 10^{7} cm^{-1}, E \sim 10^{3}\; V
cm^{-1}, m^{*} \sim 0.1 m_e$. According to our estimates, the
dimensionless probability of spin flip is equal approximately
$\sim 3.7\times 10^{-17}$. This result can be interpreted as
follows. If we assume the number of spins $N_0 \sim
[\epsilon_0^{-1}]$, where $[x]$ stands for \it entier(x) \rm, one
spin among these $N_0$ undergoes the spin flip with probability
closed to unit due to interaction considered above. If the number
of spins is $N < N_0$, the probability to find one of them
flip-flopped, is equal to $N/N_0$. The only relevant value of the
dimension of time, by means of which this process can be
characterized, is the time spent by electron within the domain
where the interaction which causes spin flips, is essential.
According to our estimates, the width of this domain ($l_0$) is of
the order of $l_0 =0.33\times 10^{-5} cm$, and the time which
takes the electron to cross this region is about $t_0\sim
l_0/v_F$. Let us assume the concentration of electron gas is equal
to $\sim 10^{17} cm^{-3}$, the cross section of the channel of the
structure in Fig.1 to be $1.0 \times 10^{-5} \times 1.0\times
10^{-5} cm$ and the length of the channel to be $100 \mu m$.  As a
result, the time of spin flip due to interaction with the surface
can be estimated as $t_0(N_0/N) \sim 0.11 sec $. This time, in
accordance with the ideology of [4] can be identified with the
spin phase relaxation time due to scattering by the surface. Now
it is clear that the time of the spin flips due to scattering by
the edges and the surface of the structure is extremely long.
Considering two mechanisms of spin flips (the scattering by
phonons and by surface or the edges) as independent, we can
calculate the total probability of spin flip as the sum of the
probabilities of these two events. As a result, the inverse spin
relaxation time is the sum of the inverse relaxation times for two
scattering mechanisms:
$(\tau^{(s)}_{\varphi})^{-1}=(\tau^{(s)}_{\varphi, ph})^{-1}+
(\tau^{(s)}_{\varphi, s})^{-1}$, where $\tau^{(s)}_{\varphi, ph},
\tau^{(s)}_{\varphi, s}$ stand for the phase relaxation times due
to scattering by the phonons
and by surface, respectively. Since $\tau^{(s)}_{\varphi, ph}\ll
\tau^{(s)}_{\varphi, s}$, the total spin relaxation time is
practically  equal to $\tau^{(s)}_{\varphi, ph}$. We can conclude
that the scattering by the surface (edges) of the structure does
not contribute essentially to the spin relaxation time and can be
neglected. Therefore, $L^{(s)}_{\varphi}$ is indeed much greater
than $L^{(e)}_{\varphi}$ and hence, the structure length can be
chosen to be
$\;$ $L_{\varphi}^{(s)}\;>\;L\;>\;L_{\varphi}^{(e)}$.\\
\section{Precession mechanisms of spin phase relaxation}

So far our assertion, $L^{(s)}_{\varphi}>>L^{(e)}_{\varphi}$ was proved to be valid for two possible mechanisms of spin phase 
relaxation: Elliot-Yafet process and the spin relaxation due to surface scattering. However, there are another possible 
causes for the destruction of 'phase memory' of the spin part of electron wave function. These are, so called 'precession 
mechanisms' of spin-relaxation, for instance, D'yakonov-Perel mechanism [14]. This mechanism is characteristic one for the 
crystals of zinc-blende structure whose point group has no inversion symmetry: in a material with bulk inversion asymmetry 
(BIA) the electron energy bands are spin split for a given direction of the wave vector $\bf k \rm$. There is also another 
very similar mechanism which is characteristic for the heterostructures and 2DEG layers, where the spin splitting may occur 
as a result of the structure inversion asymmetry (SIA) and which was first pointed out by Bychkov and Rashba [15].

Let us start with the D'yakonov-Perel mechanism. As is known, the diamond-type lattice consists of two similar face-centered 
cubic sublattices mutually penetrating each other. The zinc-blende type crystal lattices ($A^{III}B^{V}$-type, for example) 
differ from the diamond-type one in that respect, that two sublattices are not identical: the first one consists of the $A$-
type atoms, while another sublattice is of the $B$-type atoms. As a result, the point group of the zinc-blende structure does 
not involve the inversion and, as a consequence, the periodic part of the Bloch function satisfies no longer the condition 
$U_{-\bf k \rm}(\bf r \rm) =U_{\bf k \rm }(-\bf r \rm)$. Hence, a twofold degeneracy is lifted and the electron energy bands 
are spin split in these materials for a given direction of the wave vector $\bf k \rm$, even if the external magnetic field 
is not present. As a result, another mechanism of spin relaxation which was proposed for the first time in [14], becomes 
possible. We call this mechanism 'precession' mechanism of spin relaxation.

The spin splitting in $\bf k \rm$-point is equal [14,16]:
\begin{displaymath}
\Delta\varepsilon_{\bf k \rm}=\eta(\bf n \rm _{0})k^3,\;\;\;\; \bf n \rm_{0}=\bf k \rm /k,
\end{displaymath}
where $\eta(\bf n \rm_{0})=\eta _{0}\mid\bf \kappa \rm(\bf n\rm)\mid, \;\;\;\kappa_{x}=n_{0x}(n^{2}_{0y}-n^{2}_{0z})$ and the 
other components of $\bf \kappa\rm$ can be obtained by means of cyclic permutation of indices.

The spin relaxation time could be estimated as follows. Let the initial electron state characterised by $\bf k \rm$ be 
polarized along some axis, say $\bf a \rm$, which does not coinside with $\bf \kappa\rm (\bf n \rm)$ and let its spin state 
be $\sigma=+1/2$. Regarding the spin splitting $\Delta\varepsilon_{\bf k\rm}$, this state is not longer the eigenstate and as 
time passes by, it changes. It means, as it was already mentioned above, that another component with $\sigma=-1/2$ mixes up 
to the state with $\sigma=+1/2$. The coherent mixture of these two states corresponds, in the classical picture, to the spin 
precession about the $\bf \kappa \rm(\bf n\rm)$-axis with the frequency $\Omega_{\bf k \rm}=\Delta\varepsilon_{\bf k \rm}/
\hbar$.

The spin relaxation arises due to the momentum relaxation which is always present to some extent and accompanies the spin 
precession. Since $\Omega_{\bf k \rm}\tau_{\bf p \rm}<<1$, during the elapse of time between two successive scattering events 
the electron spin revolves only by a small angle $\sim \Omega_{\bf k \rm}\tau_{\bf p\rm}$. At each scattering wave vector 
$\bf k\rm$ changes randomly and as a result, $\kappa (\bf n\rm)$ changes too. It means, that the direction of spin precession 
axis changes randomly also, as the electron moves through the crystal (see Fig.3). After many scattering events the initial 
'spot'corresponding to the initial direction of precession axis spreads over unit sphere $S^2$; so, one can treat this 
process as some kind of the diffusion over $S^2$ with the angular diffusion coefficient $\cal D\rm_{s}$ of about $\sim
(\Omega_{\bf k \rm}\tau_{\bf p\rm})^2/\tau_{\bf p\rm}$.[16] 
 The time $T$ which is needed for the initial 'spot' to run over $S^2$ uniformly, can be estimated as $(\cal D\rm_{s}\em T
\rm)^{1/2}\sim4\pi$, and hence,
\begin{equation}
T\sim(4\pi)^{2}/\Omega^{2}_{\bf k \rm}\tau_{\bf p \rm}
\end{equation}
 Intuitively it is clear that $T>>\tau_{\bf p\rm}$, since one need to have many scattering events, in order the initial 
'spot' to run over entire $S^2$.

More precise calculations (see [8]) give the next formula for $\Omega_{\bf k\rm}$:
\begin{equation}
\Omega^{2}_{k}(\varepsilon)=\frac{16}{315}
\left(
\frac{\Delta}{\hbar}
\right)
\left(\frac{A^{\prime}}{\hbar^{2}/2m_{e}}
\right)
\left(
1-\frac{m^{*}}{m_{e}}
\right)
\times\frac{1}{(1+\lambda)(1+\frac{2}{3}\lambda)}
\left(
\frac{m^{*}}{m_{e}}
\right)
\left(
\frac{\varepsilon}{E_{g}}
\right)
^3,
\end{equation}
where $\lambda=\Delta/E_{g}$, $\Delta$ is the spin-orbit splitting, $E_{g}$-energy gap and the parametr $A^{\prime}$ is 
defined as follows
\begin{displaymath}
A^{\prime}=2\frac{\hbar^{2}}{m^{2}_{e}}\sum_{u}\frac{<s|P_{x}|u><u|P_{y}|z>}{E_{c}-E_{u}}.
\end{displaymath}
In the last formula $s$ is a function transformed in accordance with $\Gamma_{1}$ irreducible representation of zinc-blende 
structure space group, $u$ is the periodic part of the Bloch function and $z$ is the function which transforms under the 
tetrahedral group transformations just like $P_{z}$ atomic functions .

According to G.Fishman and G.Lampel [8], $\Omega^{2}_{k}(\varepsilon)$ for $GaAs$ is approximately equal to $10^{18}
\varepsilon^{3}$, if $\Omega$ is in rad per second and $\varepsilon$ is in $eV$.

Since we deal with mesoscopic loop structure and 2DEG, it is also important  to take into account another very similar 
mechanism of spin-splitting occurring in 2D electron gas  and which was pointed out by Yu.Bychkov and E.Rashba [15]. They 
noticed, that in heterostructures and surface layers,  there is lack of inversion asymmetry due to the existence of 
interfaces. This type of asymmetry could be called 'structure inversion asymmetry'(SIA). The corresponding spin-orbit 
Hamiltonian, according to [15] is of the form:
\begin{equation}
H_{SO}=\xi[\mbox{\boldmath $ \sigma\times {\bf{k}}]\cdot\nu$},
\end{equation}
where $\mbox{\boldmath$ \sigma$}$ are the Pauli matrices, $\mbox{\boldmath $ \nu$}$ is a unit vector perpendicular to the 
surface and $\xi $-some constant whose numerical value can be established by the cyclotron resonance data.

As in previous case, the operator $H_{SO}$ lifts the twofold spin degeneracy at $\bf k\rm\neq0$ and determines the spin-orbit 
band splitting near $\bf k\rm=0$

Just in the same way as previously, this mechanism leads to the precession of spin axis and because of $\bf k\rm$-dependence 
and electron scattering, to the diffusion of the initial 'spot' corresponding to the initial state of spin precession axis 
over entire $S^{2}$-surface. The frequency of the precession is equal to $\Omega_{{\bf {k}}\nu}=\Delta\varepsilon_{{\bf {k}} 
\nu}/\hbar$, where $\Delta\varepsilon_{{\bf {k}}\nu}$ corresponds to the spin splitting due to $H_{SO}$ (we add here 
subscript $\mbox{\boldmath$\nu$}$ in order to distinguish this mechanism of spin precession from the first one).

It is clear, that if the semiconductor, of which the structure in question is made, is of zinc-blende type, we could expect 
these two mechanisms doing simultaneously together.\\ What is less obvious and which was first pointed out by P.Pfeffer and 
W.Zawadzki [17], that there is no simple additivity of these two mechanisms: $\Delta\varepsilon_{tot}\neq\Delta\varepsilon_
{\bf k}+\Delta\varepsilon_{{\bf {k}}\nu}$ (subscript 'tot' stands for 'total"). According to [17], the $\Delta\varepsilon_
{tot}$ depends on the subtleties of semiconductor band structure and electron density in 2DEG. Now proceed to the estimates 
of the spin relaxation time for the materials of zinc-blende type, such as $GaAs$, $InSb$ and $InAs$. To the authors 
knowledge, the most reliable data are known for the first of these three materials, so let us start from the estimates for 
$GaAs$. If one starts with D'yakonov-Perel mechanism which corresponds to BIA, then one can easily estimate $\Omega_{\bf k
\rm}(GaAs)$ as to be equal $\sim3.088\times10^{11}Hz$. Taking into account SIA and its generalization proposed by Pfeffer and 
Zawadzki, supposing the electron density in 2DEG to be equal $N_{s}=10^{12}cm^{-2}$, one get from the data of [17] that 
$\Delta \varepsilon_{tot}(GaAs)$ is about $0.46 meV$. Introducing $\Omega({{\bf {k}},\mbox{\boldmath$\nu$}})=\Delta
\varepsilon_{tot}/\hbar$ and using the approach discussed above (see formula (7)) we can evaluate the spin relaxation time in 
the framework of the generalized model which includes both mechanisms. The time, according to our estimations, is 
approximately equal $T\sim 3.2\times10^{-8}$.\\ Using the data of [8], one can get for the ratios $\Omega_{\bf k\rm}(GaAs)/
\Omega_{\bf k\rm}(InSb)$ and$\Omega_{\bf k\rm}(GaAs)/\\
\Omega_{\bf k\rm}(InAs)$ the values: 0.378 and 1.58, respectively. As a consequence, $\Omega_{\bf k\rm}(InSb)\sim8.037
\times10^{11}Hz$ and $\Omega_{\bf k\rm}(InAs)\sim1.216\times10^{11}Hz$. Unfortunately, nothing is known for certain about SIA 
for these materials, as well as for its Pfeffer-Zawadzki generalization. Let us suppose, however, that the last one leads to 
the same consequences for these two materials as in previous case of gallium arsenide. Then, at $N_{s}=10^{12}cm^{-2}$ 
$\Delta\varepsilon_{tot}$ should be about twice as great as $\Delta\varepsilon_{\bf k\rm}$, and the same is valid for $\Omega
({\bf k\rm},\mbox{\boldmath$\nu$})$. As a result, we have for $T(InSb)\sim2.4\times10^{-8}s$ and for $T(InAs)\sim1.0\times10^
{-6}s$. Of course, the last results are only very rough estimates.

The natural question which is to be answered now is this: how does this spin relaxation time relate to the spin phase relaxation 
time $\tau^{(s)}_{\varphi}$ introduced above? One can relate the spin relaxation time $T$ to the phase relaxation time simply 
in the following way. Since our structure is in external magnetic field, the frequency of spin precession is the sum of 
$\Omega({{\bf {k}} ,\mbox{\boldmath$\nu$}})$ and the Zeeman frequency $\Omega_{Z}=g\mu_{B}B/\hbar$: $\Omega=\Omega({{\bf 
{k}},\mbox{\boldmath$\nu$}})+\Omega_{Z}$, where $g$-is Lande factor. This frequency is a bit different for different 
electrons, since $\Omega({{\bf {k}},\mbox{\boldmath$\nu$}})$ depends on the electron's quasi-momentum $\bf k\rm$. Thus, the 
electron scattering means that the 'oscillator' characterized by $\Omega$ is 'triggering' all the time from the frequency 
$\Omega^{\prime}({{\bf {k^{\prime}}}\mbox{\boldmath$\nu$}})+\Omega_{Z}$ to another one, $\Omega^{\prime\prime}({{\bf {k^
{\prime\prime}}}\mbox{\boldmath$\nu$}})+\Omega_{Z}$, than to $\Omega^{\prime\prime\prime}({{\bf {k^{\prime\prime\prime}}}
\mbox{\boldmath$\nu$}})+\Omega_{Z}$ and so on. Since each of these frequencies differ from one another only by some small 
value, each single scattering event leads only to small 'phase aberration'. Thus, the single 'triggering'does not yet 
introduce the irreversibility and phase destruction. However, after many scattering events not only the direction of initial 
quasi-momentum changes, but its absolute value changes too. The last one introduces the necessary element of irreversibility 
and means the destruction of 'phase memory' of the spin part of electron wave function. The complete phase destruction occurs 
after the elapse of time $\tau^{(s)}_{\varphi,PZ}\sim T\simeq (\Omega^{2}_{tot}\tau_{\bf p\rm})^{-1}$ (we add here the 
subscript 'PZ', to emphasize that we used the generalized Pfeffer-Zawadzki model).\\
Now compare the Elliot-Yafet (EY) and precession mechanisms and estimate their combined action on the spin phase relaxation. 
Note also that this kind of spin phase relaxation can be termed as environmental decoherence [1].

To this end let us make at first some comments concerning calculations of the spin phase relaxation time which is due to EY-
mechanism. In the paper [4] the next formula for $ \tau^{(s)}_{\varphi,ph}$ was derived:
\begin{displaymath}
\tau^{(s)}_{\varphi,ph}\sim(\hbar^{2}/\tau \varepsilon^{2}_{int})ln2\times tanh(\beta \varepsilon/2),
\end{displaymath}
where $\beta =(k_{B}T)^{-1}$, $\varepsilon$ - is the Zeeman splitting, $\tau$ - has the meaning of 'electron-phonon collision 
time' and $\varepsilon_{int}$ is an 'interaction energy' which characterizes an interaction between two-level quantum sub-
system and the phonon bath. In the paper [4] only the lower boundary for $\tau^{(s)}_{\varphi,ph}$ was established, which 
nevertheless turned out to be much greater than $\tau^{(e)}_{\varphi}$ (see also the final passage of Sec.2). It is possible 
to get more rigorous theory which could make more precise calculations possible. The theory could, for instance, utilize the 
spin-orbit operator of the form similar to (4), in which however, $\bigtriangledown U$ - term should be substituted by 
$\bigtriangledown V_{\bf q\rm }({\bf r\rm})$, where $V_{\bf q\rm}({\bf r \rm})$ stands for the perturbation caused by a 
single phonon. Such theory is however beyond the scope of the paper, the more so that in this eventual theory there would be 
some other parameters which in their turn are not absolutely precise. Since our aim is only to show that $\tau^{(s)}_
{\varphi,ph}>>\tau^{(e)}_{\varphi}$, we restrict our treatment rather to estimates. But in order to take into account the 
difference between $GaAs, InSb$ and $InAs$, we include into $\varepsilon_{int}$ the factor $(\bar{Z}\alpha)$, where $\bar{Z}
$ is the mean atomic number of the corresponding compound and $\alpha=e^{2}/\hbar c$. This factor is important, because the 
role of spin-orbit interaction increases as the atomic number $Z$ increases [16].

The results of our calculations are summed up in Fig. (4) and in \em Table 1\rm . In Fig. (4) the spin phase relaxation times 
which are due to EY mechanism are shown for $GaAs, InSb$ and $InAs$ as the function of external magnetic field. In $\em  
Table\rm $ $\em 1\rm$  the spin phase relaxation times, due to EY mechanism and for two chosen values of magnetic field are 
compared to the relaxation times due to precession mechanism. Obviously, one can treat these mechanisms as independent and 
hence,
\begin{displaymath}
\frac {1}{\tau^{(s)}_{\varphi}}=\frac{1}{\tau^{(s)}_{\varphi,ph}}+\frac{1}{\tau^{(s)}_{\varphi,PZ}}.
\end{displaymath}
Now it is clear that at least in accordance with our calculations, for $GaAs$ the EY-mechanism dominates for magnetic 
fields $B\leq 1T$, because $\tau^{(s)}_{\varphi,ph}$ is the shortest time. For $InSb$ one can conlude from the data of \em 
Table 1\rm , that for small magnetic field precession mechanism dominates, because $\tau^{(s)}_{\varphi,PZ}$ is the shortest 
time, while for $B\sim1T$ EY-mechanism becomes dominating.

\begin{center}
\begin{tabular}{|c|c|c|c|}\multicolumn{3}{l}{\em Table 1}\\ \hline
\em semiconductor &\multicolumn {2}{|c|}{$\tau^{(s)}_{\varphi,ph},s$} & {$\tau^{(s)}_{\varphi,PZ},s$} \\ 
\cline{2-3} 
&$B=0.1 T$ & $B=1 T$ & 

 \\ \hline

\em GaAs & $2.6\times10^{-9}$ & $2.3\times10^{-8}$ & $3.2\times10^{-8}$\\ \hline
\em InSb & $4.27\times10^{-8}$ & $1.3\times10^{-7}$ & $2.4\times10^{-8}$\\ \hline
\em InAs & $2.9\times10^{-9}$ & $2.21\times10^{-7}$ & $1.0\times10^{-6}$\\ \hline
\end{tabular}
\end{center}

It is interesting to compare our results with the experimental ones. According to the recent data obtained by D.Awschalom, 
J.Kikkawa and others [1,2], the spin decay due to environmental decoherence can exceed even 100 ns and it is in excellent 
agreement with our calculations. So, again we can conclude that the condition $L^{(s)}_{\varphi}>L>L^{(e)}_{\varphi}$ can be 
satisfied.

\section{Calculation of the transmission coefficient}

    The current $I$ through the structure considered in the Section 2, for the small applied
potential $V$, can be written as [4,18]:
\begin{equation}
I=\frac{2e}{h}\int d{\cal E}\;\int(w_{z}dk_{z}/2\pi)[f({\cal E})-f({\cal E}+eV)\;\sum_{n^{\prime},n^{\prime\prime}}
|T_{n^{\prime},n^{\prime\prime}}|^2.
\end{equation}
Here $w_{z}$ is the width of the structure in the $z$-direction, $T_{n^{\prime},n^{\prime\prime}}$
is the transmission coefficient from the state $n^{\prime}$ in the left-hand end to the state
$n^{\prime\prime}$ in the right-hand end, ${\cal E}$ and $k_{z}$
are the energy and the transverse wave vector of the electrons as they enter from the left-hand end.

The approach to calculation of the transmission coefficient $T_{n^{\prime},n^{\prime\prime}}$ was outlined in Ref.[4]; here 
for reader's convenience, we add only few comments.

Since the length $L$ of the structure is supposed to be greater than $L^{(e)}_{\varphi}$, the key idea is to devide the 
structure into sections of the length smaller than $L^{(e)}_{\varphi}$. Then one can combine these successive scatters, 
considering the transport through the states $k^{\prime}$, $k^{\prime\prime}$ as incoherent, while the transport through the 
states $\sigma^{\prime}$, $\sigma^{\prime\prime}$ as coherent because of $L<L^{(s)}_{\varphi}$. As a result, the expression 
for the transmission coefficient is of the form (see[4]):

\begin{displaymath}
T=t^{\prime}\left[I - PrP^{\prime}r^{\prime}\right]^{-1}Pt,
\end{displaymath}
where the subscripts $\sigma^{\prime}$, $\sigma^{\prime\prime}$ are dropped out.
Here $I$ is the unit matrix, $t$ is $4\times 1$ matrix describing the transmission from the left-hand end into the two 
channels,
while $t^{\prime}$ is $1\times 4$ matrix describing the
transmission from the channels into the right-hand end. Similarly, $r$ and $r^{\prime}$ are $4\times 4$ matrices
describing the reflections at the two junctions of the channels back into the channels. Matrices $P$ and $P^{\prime}$ 
describe forward
and reverse propagation of the electron wave through the channels $1$ and $2$, respectively.

The general remark which should be made, concerns the validity of Landauer-B\"uttiker formalism in this particular case. It 
is known [19] that this formalism provides a rigorous framework for the description of mesoscopic transport as long as 
transport across the structure is coherent. For noncoherent transport, however, the situation is more complicated, if there 
is a 'vertical flow' of electrons, that is the electron transitions from one energy to another. In that sence our case is 
rather intermediate one: the transport through the states $\sigma^{\prime}$, $\sigma^{\prime\prime}$ is coherent, while 
through the states $k^{\prime}$, $k^{\prime\prime}$ is incoherent. Luckily, sometimes even if 'vertical flow' is pressent, it 
can be neglected (see[19,p.111]) provided transmission functions are approximately constant over the energy range where 
transport occurs:
\begin{displaymath}
(\mu_{1}+\mu _{2})+(n\times k_{B}T)<<\varepsilon_{c}\;,
\end{displaymath}
where $1\leq n \leq 5$, $\varepsilon_{c}$ is the correlation energy.\\
To estimate correlation energy, one can use simple relation: $\varepsilon_{c}\sim \hbar/\tau^{(e)}_{\varphi}$. Since $\tau^
{(s)}_{\varphi}>>\tau^{(e)}_{\varphi}\sim 10^{-12}-10^{-13}$, correlation energy is about 0.6-6 meV. So, we assume the 
voltage $V$ applied to the structure to be sufficiently small, in order to satisfy the condition above.

Since we assume the length of the structure $L>L^{(s)}_{\varphi}$, there are no spin flips in two channels considered and 
hence, among the 16 matrix elements of $r$-matrix (as well as $r^{\prime}$) 8 entries are equal to zero.

In order to construct $P$ and $P^{\prime}$, it is necessary to note that the spin parts of the wave functions acquire the 
phase
factors due to Larmor spin precession about \bf B \rm -axis. Since magnetic field in the channels are different, these phase 
factors
are also different.

One can treat the states 'spin up' and 'spin down' as the two opposite points on a unit sphere ${\cal S}^2$ which can be 
transformed
one into another under rotation by an angle $\varphi =\pm\pi$ about some axis \bf a\rm. Introduce also formally  \bf b\rm -
axis which
is a unit vector of the precession axis: $+\bf b$
\rm corresponds to the electron propagation from $x  =  0$ to $x  =  L$  while $-\bf b$\rm $\;$
corresponds to reverse propagation, and $\theta_{1}$ and $\theta_{2}$ are the phase acquired by spin part of the wave 
functions in the
channels $1$ and $2$, respectively.  Then the matrix
elements describing the phase shifts in the two channels can be written as:
\begin{eqnarray}
P_{\pm 1} = \exp(\pm i\varphi_a)\exp(i\theta_{1,b}),\;
P_{\pm 1}^{\prime} = \exp(\pm i\varphi_a)\exp(-i\theta_{1,b}),\\
P_{\pm 2} = \exp(\pm i\varphi_a)\exp(i\theta_{2,b}),\; P_{\pm
2}^{\prime} = \exp(\pm i\varphi_a)\exp(-i\theta_{2,b}).
\end{eqnarray}
\\
The idea of (8)-(9) is to express the elements of the matrices $P$, $P^{\prime}$ as the two rotations about two independent 
axis.
Then, these objects are nothing else but the unitary quaternions [20].
As is known [20], any quaternion can be written in the form $q =c_0 + i_1c_1 + i_2c_2 + i_3c_3 =\sum^{3}_{\alpha =0} i_
{\alpha}c_{\alpha}$,
where $i_0=1$ and $i_1^{2}=i_2^{2}= i_3^{2}= i_1i_2i_3=-1$. However, it is possible also to define, for instance, $i_1, i_2$ 
as
\begin{displaymath} i_1 =
\left(
\begin{array}{cc}
0 & 1 \\
-1 & 0
\end{array}
\right), i_2 = \left(
\begin{array}{cc}
0 & i \\
i & 0
\end{array}
\right),
\end{displaymath}
where $i$ is the ordinary complex square root of $-1$, thus forcing
\begin{displaymath} i_3 = i_1i_2 = \left(
\begin{array}{cc}
i & 0 \\
0 & -i
\end{array}
\right). \end{displaymath} If these three matrices are multiplied
by $-i$, one obtains Pauli spin matrices. Thus, the quaternion $q$
could have been identified with the complex 2-by-2 matrix
\begin{displaymath} \left (
\begin{array}{cc}
c_0 + ic_3& c_1 + ic_2\\
-c_1 + ic_2 & c_0 -ic_3\end{array}\right) =\left ( \begin{array}{cc}
u & v\\
-v^{*}& u^{*}\end{array}\right ),\end{displaymath}

where $u$ and $v$ are complex numbers with complex conjugates $u^{*}$ and $v^{*}$. Replacing $0,1$ and  $i$ in these complex
matrices by \begin{displaymath}\left (
\begin{array}{cc}
0 & 0\\
0 & 0 \end{array}\right ), \left ( \begin{array}{cc} 1 & 0\\ 0 & 1\end{array}\right ), \left ( \begin{array}{cc} 0 & 1\\-1 & 
0 \end{array}\right ),
\end{displaymath}
respectively, one can obtain a representation of quaternions as 4-by-4 matrices.

Since two channels $1$
and $2$ are supposed to be isolated, in this way the matrices
$P$ and $P^{'}$ can be represented as the diagonal $4\times 4$-matrices with the diagonal elements defined by (8)-(9).\\
After a great deal of algebra (see Ref.[4] ), we have:
\begin{eqnarray*}
|T|^{2} = |a_{1}|^{2} + |a_{2}|^{2} + |a_{3}|^2 + |a_{4}|^2 + (a_{1}^{\ast}a_{3} + a_{1}a_{3}^{\ast}
+ a_{2}^{\ast}a_{4} + a_{2}a_{4}^{\ast}) + (a_{1}^{\ast}a_{2} \\
+ a_{1}a_{2}^{\ast} + a_{2}^{\ast}a_{3}+ a_{2}a_{3}^{\ast} + a_{3}^{\ast}a_{4} + a_{3}a_{4}^{\ast})cos \Delta\theta,
 \Delta\theta = \theta_{1}- \theta_{2},
\end{eqnarray*}
where $a_{i}, (i=1,2,3,4)$ do not depend on $\theta_{1}$,$\theta_{2}$ and are the complicated functions
of $r_{ij}$, $r_{ij}^{\prime}$, $t_{i}$, $t_{i}^{\prime}$.\\ Note, that the cosine dependence of the transmission coefficient 
on phase difference in the last expression is the direct consequence of the quaternion representation of the propagation 
matrices, $P$ and $P^{\prime}$.

It is interesting to note that the same cosine dependence on phase defference was obtained in the experiments with two 
interfering neutron beams [21]. These results are now considered as the direct verification of the $4\pi$-symmetry of 
spinors. Since Dirac's equation can be written in quaternion representation (see, for instance, [22]), we can conclude that 
our result is in total agreement with $4\pi$-symmetry of spinors.

\section{Calculation of the phase shift}

Now let us suppose the mesoscopic structure under consideration is made od the zinc-blende type semiconductor and proceed to 
the calculation of the phase shift acquired by the spin part of the electron wave function.

    Consider  the non-relativistic motion of the particle (electron) with the spin $ |s| = 1/2$ in a two-
component magnetic field:$\;$ $\bf B\rm = \bf B_{0} \rm + \bf B_{1}$ \rm , $\bf B_{0} \rm = (0,B_{0},0)$,
and $\bf B_{1}\rm = (0,0,B_{1})$, where $B_{1}$ is an additional uniform magnetic field in one of the channels of the 
structure.
The spin part of electron wave
function can be considered as a two-component vector defined by the pair of functions $\chi(|\uparrow>)$ and $\chi(|
\downarrow>)$
which stand for the probability amplitudes of the two possible orientations of spin. The spin operator
$\hat{\sigma}(\sigma_{x},\sigma_{y},\sigma_{z})$ is defined in terms of Pauli matrices:
\begin{displaymath}
\sigma_{x}=\left(
\begin{array}{cc}
0 & 1\\
1 & 0
\end{array}
\right),
\sigma_{y}=\left(
\begin{array}{cr}
0 & -i\\
i & 0
\end{array}
\right),
\sigma_{z}=\left(
\begin{array}{cr}
1 & 0\\
0 & -1
\end{array}
\right)
\end{displaymath}
Thus,we can treat the mean
value of the magnetic moment of the electron moving within the channels of mesoscopic structure as the classical
quantity $\bf P\rm =<\sigma>$, its evolution under magnetic field being defined by the equation:
\begin{displaymath}
\frac{d\bf P}{dt} = \gamma^{*}\left[\bf P,B\rm \right],
\end{displaymath}
where $\gamma^{*} = e/mc$ is the electron gyromagnetic constant.

In other words, the vector $\bf P$ can be treated as classical magnetic top and, if this classical top
having the initial orientation $\bf P_{0}\rm = (P^{0}_{x},P^{0}_{y},P^{0}_{z})$ enters magnetic
field $\bf B\rm = (B_{x},B_{y},B_{z})$, it begins to precess about magnetic field with the frequency
$\Omega = \gamma^{*} B$, where $B = \sqrt{B_{x}^{2}+B_{y}^{2}+B_{z}^{2}}$.

It is interesting to note that despite its purely quantum
character, the spin of the particle during its movement in
external fields often can be treated classically. The accuracy of
such treatment can be estimated by means of Heisenberg uncertainty
relation, since classical treatment is possible if one can neglect
the commutator
 $[\bf r \rm ,\bf p\rm ]$ where $\bf p$ \rm is the particle momentum operator. So, the measure of accuracy of the classical
approximation is
$|\Delta p|/p$. $\Delta p$ in our case can be estimated as $\sim m\Delta v = m(v^{2}/l_B)\Delta t$, where $l_B = \sqrt{\hbar 
c/|e|B}$
is the magnetic
length and $\Delta t \sim 2\pi/\omega_c, \omega_c = |e|B/mc$ is the cyclotron frequency, while 
$|\Delta p|/p \sim 2\pi mv_F \sqrt{mc}/\sqrt{\hbar |e|B}$. Assuming $v_F\sim 3\times 10^{7} cm s^{-1}$ and $B\sim 0.1 T$, we 
have
$|\Delta p|/p   \approx 1.26\times 10^{-10}$. Therefore, indeed to a good approximation, we can treat the evolution of vector 
$\bf P$
as the evolution of the classical magnetic top under external magnetic field.

Let us introduce now the phase of precessing spin by means of the formula
\begin{displaymath}
\theta(v,x) = \mu_{B}g/\hbar \int_{0}^{x} B(v,x)dt = \gamma^{*}/v\int_{0}^{x} B dx^{\prime}.
\end{displaymath}
 Since magnetic fields, $\bf B_{0}\rm$ and $\bf B_{1}\rm$ are uniform, the calculation of the phase shift $\Delta\theta$ can 
easily be done. 

Moreover, it is clear that under certain conditions
including appropriate structure length $L$, electron velocity and the values
of magnetic fields $B_{0},\;B_{1}$,$\;$ the phase shift $\Delta \theta = \theta_{2} - \theta _{1}$
can be multiple of $\pi$. Indeed,
\begin{displaymath}
\Delta\theta = \theta_{2} -\theta_{1} = n\pi = (\gamma L/v)(\sqrt{B^{2}_0 + B^{2}_1}- B_0 )\;\;  ,n=1,2...
\end{displaymath}
If the values of $B_{1}, L, v,n$ are given, the value of $B_{0}$ which is needed for the $\Delta\theta$
to be equal of multiple of $\pi$ can be easily calculated:
\begin{equation}
B_0 = 
\left|
\frac{\gamma^{*} L}{2n\pi v}B^{2}_1 - \frac{n\pi v}{2\gamma^{*} L}
\right |.
\end{equation}
Hence, changing the external magnetic field $B_{0}$, one can
change the phase shift and the quantum interference from
constructive to destructive one and back. Also it is seen that
$\Delta\theta =\theta_{2}-\theta_{1}=f(B_{0}$,$B_{1},v)$ is the
function of $B_{0}$, $B_{1}$,$v$. That is, the phase shift
generally speaking is different for the electrons with different
velocities. At first sight, this makes matters worse, because it
means that the 'interference pattern' should be blurred. One
should remember, however, that the temperature is considered to be
sufficiently low. It means that most of the electrons carrying current are on the Fermi surface, that is the electron 
distribution function $f(\varepsilon)=\chi (\varepsilon_{F}-\varepsilon)$ and $v=v_{F}$, where $\chi {(...)}$ is the 
Heaviside step-like function, $\varepsilon_{F}, v_{F}$ are the Fermi energy and Fermi velocity, respectively. Someone can be 
temptated to substitute $v$ in the last formula by drift velocity. But the condition $L^{(s)}_{\varphi}>L>L^{(e)}_{\varphi}$ 
does not imply that electrons undergo so many collisions that the drift velocity arises. Remember,  
that the concept of drift velocity is relevant to macroscopic
samples, where electrons undergo a great many collisions under
which (and an external electric field) the drift velocity can only
be formed. Here instead, we have mesoscopic structure where
electrons suffer only a few collisions after which the phase
coherence of the orbital part of the electron wave functions in
the two arms of the structure is destroyed.

As a result, we can substitute in (13) $v$ by $v_{F}$ assuming that most of the electrons carrying current are on the Fermi 
surface and a few collisions which they undergo during their movement withing the channels of the structure do not change 
essentially their flight time which is still approximately equal to $\sim L/v_{F}$.\\
So, the calculation by means of (10) taking into account the expression for $|T|^{2}$, now can easily be done and we have:
\begin{displaymath}
I=(2e/h)K(A+Dcos\Delta\theta(v_{F})),
\end{displaymath}
where $K,A,D$ are the coefficients dependent on the peculiarities of the structure. Now it is clear that changing $B_{0}$ one 
can approach very deep modulation of the conductance and since $A\sim D$, the 'contrast' of the 'interference pattern' is 
defined only by the ratio $\sqrt\frac{\varepsilon_{F}-k_{B}T}{\varepsilon_{F}}$.\\
So, we conclude that if the structure length $L$ is chosen to be $L^{(s)}_{\varphi}>L>L^{(e)}_{\varphi}$, it is indeed 
possible to 'wash out' the quantum interference related to phase coherence of the 'orbital part' of electron wave function, 
retaining at the same time that one related to the phase coherence of the spin part. Moreover, we can expect this 
'interference pattern' and corresponding current (or conductance) modulation to be strong enough in order to be observed.

It is also interesting to note that current (and conductance) oscillations generally speaking, are $\em not\rm$ periodic with 
respect to $B_{0}$, the magnetic field by means of which these oscillations are controlled. This is due to the fact, that 
while the second term in (13) is linearly proportional to $n$, the first one is inversely proportional to it. Another 
interesting feature of (13) is the quadratic dependence of $B_{0}$ on the static field $B_{1}$. By means of this formula one 
can easily calculate $\Delta B_{0}$, the changing in magnetic field $B_{0}$ which is needed to change $\Delta\theta$,for 
example, from $\pi$ to $2\pi$. The corresponding data for three zinc-blende type semiconductors, two chosen values of $B_{1}
$ and $L=1.5\times10^{-2}cm$ are presented in \em Table 2\rm .\\

\begin{center}
\begin{tabular}{|c|c|c|} 
\multicolumn{3}{l} {\em Table 2} \\ \hline
\em semiconductor &\multicolumn {2}{|c|}{$\Delta B_{0},T$}\\ \cline{2-3}
& $B_{1}=0.1,T$ & $B_{1}=0.5,T$ \\ \hline
\em GaAs & 0.002 & 0.91 \\ \hline
\em InSb & 3.54 & 88.68 \\ \hline
\em InAs & 1.07 & 26.75 \\ \hline
\end{tabular}
\end{center}

Keeping in mind the possibility of experimental verification of the theory presented in the paper, one can conclude from the 
\em Table 2\rm , that some materials and some values of magnetic field $B_{1}$ are more sutable than the others. Perhaps 
$GaAs$ is the best material for that purpose, while if $B_{1}>0.2T$ such experiment for $InSb$ and $InAs$ becomes rather 
impossible.

\section{ Conclusion}

    A simple theory of the quantum interference due to Larmor precession of an electron spin in a loop structure is presented 
in this paper. We investigate different mechanisms of environmental decoherence, such as edge scattering, Elliot-Yafet and 
precession mechanisms of spin relaxation, as well as their influence on the quantum spin interference in such structure. It 
turns out, that the time of spin phase relaxation due to edge scattering is very long and this mechanism can be neglected, 
while the other two are essential. The EY - and precession mechanisms thus determine the spin phase relaxation. As it is 
shown, even if EY - and precession mechanisms do together, it is still possible nevertheless, to satisfy the condition $L^
{(s)}_{\varphi}>L>L^{(e)}_{\varphi}$. The last one determines the 'spin ballistic' transport in the structure in question, 
that is, the phase relaxation length $L^{(s)}_{\varphi}$ of the spin part of the electron wave function is assumed to be 
greater than the microstructure length.
If in one of the microstructure's arms
there is an additional magnetic field, the spin wave function
acquires a phase shift due to additional spin precession about
that field.

 Now if we suppose the
microstructure length is chosen to be greater than the
$L_{\varphi}^{(e)}$, it is possible to 'wash out' the quantum
interference related to phase coherence of the 'orbital' part of
the wave function retaining at the same time that related to the
phase coherence of the spin part and hence, reveal the
corresponding conductance oscillations.

Changing the external magnetic field, one can change the 'interference pattern', that is,to control the conductance 
modulation. We have shown that the strong conductance modulation can be achieved in this way.

\section{Acknowledgements}

    One of us (I.T.) is greatly acknowledged to Prof. Robin Nicholas, University of Oxford, UK
for discussing the topics touched upon in the paper and his profound comments.

\newpage

\bf Fig.1 \rm A sketch of a two-channel semiconductor mesoscopic structure with an additional maghetic field (\it 1\rm ) 
accross one of the channels.
On the upper panel $t, t^{\prime}, r, r^{\prime}$ indicate the transmission and reflection matrices at the two junctions $x 
\le 0, x \ge L$;
$P, P^{\prime}$ stand for the propagation matrices in the middle region ($0 \le x \le L$); \it 2\rm : the external magnetic 
field $B_0$.\\

\bf Fig.2 \rm Airy function evaluated by means of the parameters assumed (see the text).\\
$\it 1\rm - Ai(z)=Ai(\alpha z_{0}(z-1))$,$\;\;$
$\it 2\rm - F(z)=(Ai(\alpha z_{0}(z-1)))^{2}$, 
where:\\
$\alpha =
\left(
\frac{2m^{*}eE}{\hbar^{2}}+\frac {2eB_{0}k_{x}}{\hbar c}
\right)^{\frac{1}{3}}$ ,$\; \; $ $z_{0}=\frac{\varepsilon(k_{x})-\frac{\hbar^{2}k^{2}_{x}}{2m^{*}}}{eE+\frac{\hbar eB_{0}k_
{x}}{m^{*}c}}$ ,
$\varepsilon(k_{x})-\frac{\hbar^{2}k_{x}^{2}}{2m^{*}}=
\left(
\frac {\hbar^{2}}{2m^{*}}
\right)^{\frac {1}{3}}
\left[
\frac {9\pi}{8}
\left(
eE+\frac{\hbar eB_{0}k_{x}}{m^{*}c}
\right)
\right]^{\frac{2}{3}}.$\\

\bf Fig.3 \rm Spin phase relaxation time due to Elliot-Yafet mechanism versus external 
magnetic field for three zinc-blende type semiconductors.\\

\bf Fig.4 \rm Schematic representation of the precession mechnism of spin relaxation:
\it 1 - \rm $S^{2}$-sphere; \it 2 - \rm a point on the $S^{2}$-sphere corresponds the the 
initial position of the spin precession axis; \it 3 - \rm shaded cirlces correspond to the sequential positions of the spin 
precession axis whose direction changes randomly due to collisions.


\newpage

\begin{thebibliography}{25}
\bibitem{} D.D. Awschalom, J.M. Kikkawa, Physics Today \bf 52 \rm , 33 (1999)
\bibitem{} J.M. Kikkawa, J.A. Gupta, I. Malajovich, D.D. Awschalom, Physica E \bf 9 \rm , 194 (2001)
\bibitem{} S. Washburn, R.A. Webb, Adv.in Phys. \bf 35 \rm , 375 (1986)
\bibitem{} I.Tralle, J Phys: Condens. Matter \bf 11 \rm , 8239  (1999)
\bibitem{} Yu.I.Manin,\em {Mathematics and Physics} \rm (Ser. Progress in Physics, Ed. by A.Jaffe and D. Ruelle) Birkh
\"auser, Boston, Stuttgart, 1981 (p.55) 
\bibitem{} D. H\"agle, M. Oestreich, W.W. R\"ule, N. Nestle, K.Eberl Appl Phys Lett \bf 73 \rm , 1580 (1998)
\bibitem{} K.K. Choi, D.C. Tsui and K. Alavi, Phys Rev B \bf 36 \rm ,  7751 (1987)
\bibitem{} G. Fishman, G. Lampel, Phys Rev B, \bf 16 \rm , 820 (1977)
\bibitem{} F.J. Dyson, Phys Rev \bf 98 \rm ,  349 (1955)
\bibitem{} D. Bohm, \em {Quantum Theory} \rm, (NY: Prentice-Hall Inc. 1952)
\bibitem{} V.N. Lisin, B.M. Khabibulin, Fiz Tverd Tel (Sov. Solid State Phys) \bf 17 \rm , 1600 (1975)
\bibitem{} O.V. Kibis, JETP Lett \bf 66 \rm , 588  (1997)
\bibitem{} M. Abramowitz, I.A. Stegun, eds.\em { Handbook of Mathematical Functions} \rm , ( Dover Publ. Inc., 1965)
\bibitem{} M.I.D'yakonov and V.I.Perel', Fiz. Tverd. Tela, \bf 13\rm,3851 (1981)[Sov. Phys.-Solid State \bf 13\rm, 3023 
(1972)]
\bibitem{} Yu.A.Bychkov, E.I.Rashba, J.Phys C, \bf 17\rm, 6039 (1984)
\bibitem{} V.F.Gantmacher, I.B.Levinson, \em {Carrier scattering in metals and semicoductors}\rm , North-Holland, NY-Tokyo, 
1987
\bibitem{} P.Pfeffer, W.Zawadzki, Phys Rev B, \bf 52\rm, R14 332 (1995)
\bibitem{} S. Datta and S. Bandyopadhyay, Phys Rev Lett \bf 58 \rm , 717 (1987)
\bibitem{} S. Datta, \em {Electronic Transport in Mesoscopic Systems},\rm (Cambridge: Cambr, Univ. Press 1995)
\bibitem{} G. Casanova, \em {L'algebre vectorielle} \rm , ( Paris: Presses Universitaires de France 1976 )
\bibitem{} H. Rauch, Europhysics News, \bf 28 \rm, 10 (1997)
\bibitem{} J. Lambek, The Mathematical Intelligencer, \bf 17\rm, 7 (1995)

\end{thebibliography}
\end{document}